\documentclass[pdflatex, sn-nature, iicol]{sn-jnl}
\usepackage{amsmath,amssymb,amsfonts}
\usepackage{dsfont}
\usepackage{pdflscape}
\usepackage{physics}
\usepackage[separate-uncertainty=true,multi-part-units=single]{siunitx}
\usepackage[all]{hypcap}

\usepackage[sort&compress,numbers]{natbib}
\usepackage{doi}

\newcommand\mybar{\kern-2pt\rule[-.8pt]{.8pt}{7pt}\kern1pt}

\begin{document}

\title{Parity-dependent state transfer for direct entanglement generation}

\author[1,2]{F.~A.~Roy}\equalcont{}
\author[1,3]{J.~H.~Romeiro}
\equalcont{
    These authors contributed equally. Correspondence should be addressed to:
    
    \href{mailto:federico.roy@wmi.badw.de}{federico.roy@wmi.badw.de}; \href{mailto:joao.romeiro@wmi.badw.de}{joao.romeiro@wmi.badw.de}.
    }
\author[1,3]{L.~Koch}
\author[1,3]{I.~Tsitsilin}
\author[1,3]{J.~Schirk}
\author[1,3]{N.~J.~Glaser}
\author[1,3]{N.~Bruckmoser}
\author[1,3]{M.~Singh}
\author[1,3]{F.~X.~Haslbeck}
\author[1,3]{G.~B.~P.~Huber}
\author[1,3]{G.~Krylov}
\author[1]{A.~Marx}
\author[1,3]{F.~Pfeiffer}
\author[1,3]{C.~M.~F.~Schneider}
\author[1,4]{C.~Schweizer}
\author[1,3]{F.~Wallner}
\author[1,3]{D.~Bunch}
\author[1,3]{L.~Richard}
\author[1,3]{L.~Södergren}
\author[1,3]{K.~Liegener}
\author[1,3]{M.~Werninghaus}
\author[1,3,5]{S.~Filipp}

\affil[1]{Walther-Meißner-Institut, Bayerische Akademie der Wissenschaften, 85748 Garching, Germany}
\affil[2]{Theoretical Physics, Saarland University, 66123 Saarbrücken, Germany}
\affil[3]{Technical University of Munich, TUM School of Natural Sciences, Department of Physics, Garching 85748, Germany}
\affil[4]{Fakultät für Physik, Ludwig-Maximilians-Universität München, Schellingstraße 4, 80799 München, Germany}
\affil[5]{Munich Center for Quantum Science and Technology (MCQST), Schellingstraße 4, 80799 München, Germany}

\date{\today}

\abstract{
As quantum information technologies advance, challenges in scaling and connectivity persist, particularly the need for long-range qubit connectivity and efficient entanglement generation.
Perfect State Transfer enables time-optimal state transfer between distant qubits using only nearest-neighbor couplings, enhancing device connectivity. 
Moreover, the transfer protocol results in effective parity-dependent non-local interactions, extending its utility to entanglement generation.
Here, we experimentally demonstrate Perfect State Transfer and multi-qubit entanglement generation on a chain of six superconducting transmon qubits with tunable couplers, controlled via parametric drives. 
By simultaneously activating and engineering all couplings, we implement the transfer for up to six qubits, verifying single-excitation dynamics for different initial states.
Extending the protocol to multiple excitations, we confirm its parity-dependent nature, where excitation number controls the phase of the transferred state.
Finally, leveraging this property, we prepare a Greenberger-Horne-Zeilinger state using a single transfer operation, showcasing the potential of Perfect State Transfer for efficient entanglement generation.
}
\maketitle

\section*{Introduction}

Quantum information technologies have evolved significantly in the past 20 years and are now at the verge of demonstrating useful applications of quantum computing~\cite{Acharya2023,Kim2023,Bluvstein2023,Feng2023,Iqbal2024}.
Nonetheless, many technologies are facing limitations in the scalability of their platform.
In particular, efficiently connecting distant qubits within the same processing unit, and even between different processing units, poses a significant challenge for the generation of highly entangled states. 
Qubit shuttling has proven to be a successful solution to the problem of connectivity for several different platforms~\cite{Mills2019, Yoneda2021, Pino2021, Bluvstein2022, Burton2023, Kuenne2024}.
Nevertheless, some platforms feature qubits that are static in nature, as is the case for solid-state technologies such as superconducting qubits, making this scheme impossible to use.
In this scenario, information has to be moved via a series of swap operations between adjacent qubits until it reaches its destination.
As a result, entanglement generation is generally limited and requires sequences of two-qubit and single-qubit gates to achieve the desired state.
Alternatively, there have been proposals and demonstrations having multiple qubits coupled to a common element to increase connectivity and to generate many-body entanglement.
These protocols are either designed for a limited number of qubits embedded in low-connectivity architectures~\cite{Gu2021, Zhang2022, Baker2022, Kim2022, Warren2023}, or require specially designed architectures where all qubits involved must share a common coupling element ~\cite{Mezzacapo2014, Song2017, Hazra2021, Lu2022, Glaser2023}.
Consequently, both approaches face limitations in further scaling up many-body entanglement and in coupling genuinely distant qubits.

In this context, Perfect State Transfer (PST) provides an alternative approach to efficiently couple and entangle multiple qubits.
In PST, an arbitrary number of qubits is assumed to be coupled along a linear chain with no extra elements, making it a more viable technique for scalable architectures.
By activating all couplings in the chain and controlling the strength of the time-independent couplings, quantum states are time-optimally transferred between the qubits on either end of the chain~\cite{Cook1979, Peres1985, Christandl2004, Nikolopoulos2004, Albanese2004, Yung2004, Yung2005, Shi2005, Karbach2005, Petrosyan2006, Vinet2012, Vinet2012_Krawtchouk, Nikolopoulos2014, Loft2016}.
The same configuration of coupling strengths also result in state transfers between all pairs of mirror-symmetric qubits, i.e. equidistant from the center of the chain.
In fact, when considering multiple excitations, the phase of the transferred state will depend on the parity of excitations present between the initial and final transfer locations, as described in detail in~\cite{Naegele2022}.
This gives rise to multi-qubit interaction terms that can be utilised for efficient many-body entanglement creation, as well as for implementing multi-qubit gates~\cite{Clark2005, Kay2010, Brougham2011, Naegele2022}.
Compared to previous approaches, PST yields non-local connectivity between qubits and provides a method to efficiently generate entanglement without the need for modifications to the hardware.
However, experimental work so far has been limited to transfers with excitations only present in the ends of the chain~\cite{Madi1997, Zhang2005, Alvarez2010, Bellec2012, Perez-Leija2013, Chapman2016, Signoles2017, Li2018, Tian2020, Burkhart2021, Karamlou2022, Zhang2023, Xiang2024}.

Here, we experimentally demonstrate the properties of PST in the presence of multiple excitations throughout the chain.
Using superconducting transmon qubits~\cite{Koch2007} and parametrically driven tunable couplers~\cite{Mckay2016}, we implement and control the required couplings and perform the operation for different lengths of chains and prepared initial states.
Our results show that excitations transfer from each initial qubit to its mirror-symmetric counterpart. 
By transferring a superposition state, we directly observe the dependence of the final phase to the number parity of excitations, as predicted by theory.
Harnessing these properties, we generate a three-qubit Greenberger-Horne-Zeilinger~(GHZ) state with a single PST operation.
Finally, using the theory of graph states, we show how this method can be generalised to larger qubit numbers, showcasing the usefulness of the PST protocol for scalable quantum hardware.

\begin{figure*}[t]
\centering
\includegraphics{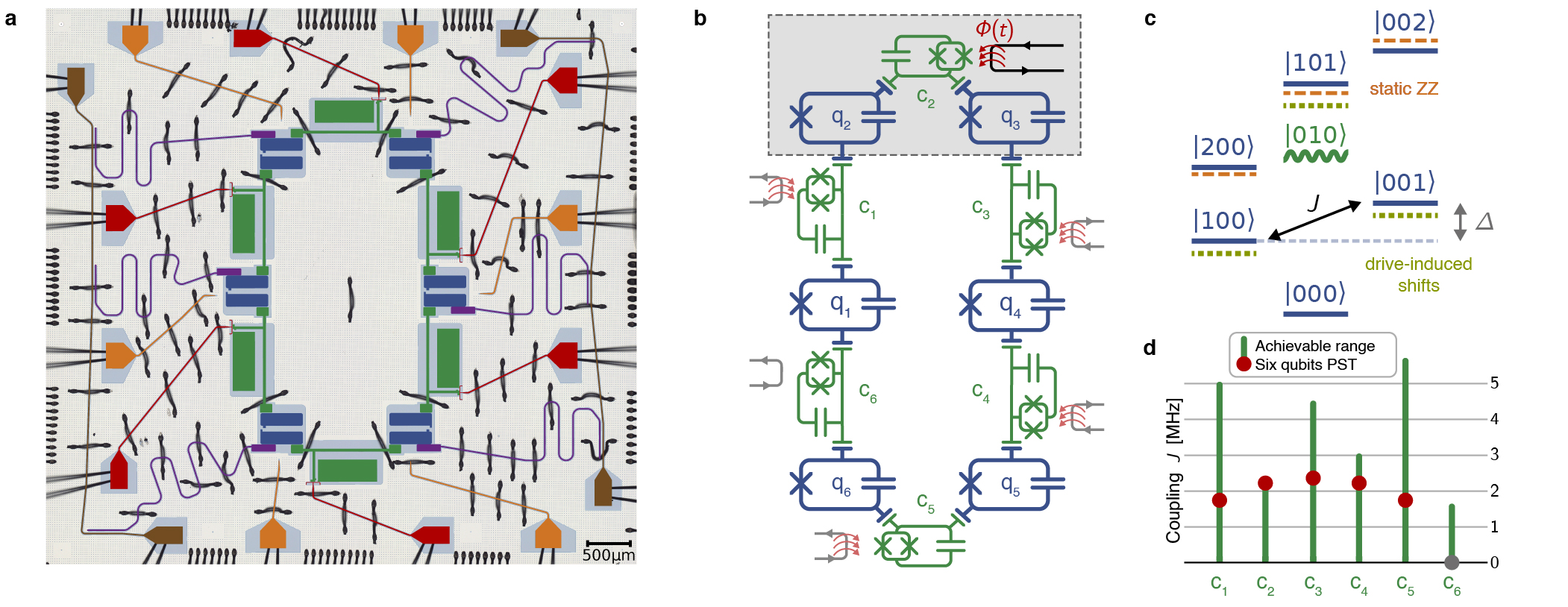}
\caption{
    \label{fig:device}
    \textbf{Device and parametric coupling.}
    \textbf{a}~False-colour image of the six-qubits device.
    Fixed-frequency transmon qubits (blue) are arranged in a ring with tunable transmon couplers (green) between them to mediate couplings.
    Individual coupler flux lines (red) and qubit drive lines (orange) enable full control of the system.
    Individual readout resonators (purple) coupled to feed lines (brown) on either side of the chip are used to readout the qubits.
    Wire bonds (black) connect all ground planes and the PCB lines to lines on the chip.
    \textbf{b}~Circuit diagram representing the device.
    All qubits (blue) couple capacitively to the couplers (green) on either side.
    The couplers are tuned in frequency via the flux $\phi(t)$ (red) threading through their respective SQUID loop.
    Parametric drives are used to activate and control effective couplings between qubits.
    \textbf{c}~Schematic energy level diagram of two neighbouring qubits (blue) and the coupler (green). The states are labelled as $\ket{\text{q}\text{c}\text{q}}$.
    A parametric flux drive modulates the coupler at the difference frequency of the qubits $\Delta$ (grey arrow) activating the effective coupling $J$ (black arrow).
    Yellow dashed lines represent the shifted levels induced by the drive.
    ZZ shifts due to the interaction of levels with two excitations are shown in orange. 
    \textbf{d}~Effective coupling strengths between all neighbouring qubit pairs on the device. 
    Green lines indicate the full range of coupling strength achievable by varying the amplitude of the parametric drive of each coupler.
    Red dots show the choice of coupling following the PST solution for $N=6$. 
    Coupler $\text{c}_6$ is not modulated (grey dot), turning off the interaction on this edge of the ring.
}
\end{figure*}

\section*{Results}

\subsection*{Device Description}
Experiments are carried out on a superconducting device hosting six fixed-frequency transmon qubits, $(\text{q}_i)$, each coupled in a ring layout to their two nearest neighbours via tunable couplers, $(\text{c}_j)$, shown in Fig.~\ref{fig:device}\textbf{a}.
All qubits have individual drive lines and readout resonators for single-qubit control and measurement, respectively.
The Hamiltonian of this system is given by
\begin{equation}
    \label{eq:Hamiltonian_system}
	\begin{split}
	\hat{H}/\hbar   = 
    &\sum_{1\leq i\leq6} \omega_{\text{q}_i} \hat{a}_{\text{q}_i}^\dagger\hat{a}_{\text{q}_i} + \frac{\alpha_{\text{q}_i}}{2} \hat{a}_{\text{q}_i}^\dagger\hat{a}_{\text{q}_i}^\dagger\hat{a}_{\text{q}_i}\hat{a}_{\text{q}_i}\\
    + &\sum_{1\leq j\leq6} \omega_{\text{c}_j}(\phi_{\text{c}_j}) \hat{a}_{\text{c}_j}^\dagger\hat{a}_{\text{c}_j} + \frac{\alpha_{\text{c}_j}}{2} \hat{a}_{\text{c}_j}^\dagger\hat{a}_{\text{c}_j}^\dagger\hat{a}_{\text{c}_j}\hat{a}_{\text{c}_j}\\
    +&\sum_{\{i,j\}} \frac{g_{ij}}{2}(\hat{a}_{\text{q}_i}^\dagger-\hat{a}_{\text{q}_i})(\hat{a}_{\text{c}_j}^\dagger-\hat{a}_{\text{c}_j}),\\
    \end{split}
\end{equation}
where $a$ and $a^\dag$ are the creation and annihilation operators for qubits ($\text{q}_i$) and couplers ($\text{c}_j$), with frequencies and anharmonicities given by $\omega_{\text{q}_i/\text{c}_j}$ and $\alpha_{\text{q}_i/\text{c}_j}$, respectively.
The strength of static couplings between qubits and their neighbouring couplers is given by $g_{ij}$.
The coupler frequencies can be individually tuned by applying external magnetic fields $\phi_{\text{c}_j}$. 
Qubit populations are measured through dispersive readout. Measurement errors are characterised using an assignment matrix and mitigated by applying matrix inversion~\cite{Bravyi2021}.
A circuit diagram representing all qubits and couplers in the device is shown in Fig.~\ref{fig:device}\textbf{b} and device parameters are given in~\ref{tab:device}.

Local qubit-qubit interactions are activated via parametric drives on the respective tunable couplers.
By applying the external flux $\phi_{\text{c}_j}(t)~= \phi_\text{dc} + A_j\cos(\Delta_j t / k)$, we modulate the coupler at the $k$-th harmonic of the difference frequency $\Delta_j$ between adjacent qubits, generating an effective interaction between them~\cite{Mckay2016}, as shown in Fig.~\ref{fig:device}\textbf{c}.
Since the couplers are in the dispersive regime ($|\omega_{\text{c}_j} - \omega_{\text{q}_i}| \gg g_{ij}$) they can be decoupled from the qubit dynamics by performing a time-dependent Schrieffer-Wolff transformation~\cite{Roth2017}.
Then, truncating the qubit states to the computational subspace yields the effective system Hamiltonian
\begin{equation}
    \label{eq:H_eff}
    \begin{split}
    \hat{H}_{\text{eff}} / \hbar = \sum_{1\leq i\leq6}\big[&J_i(\hat{\sigma}_i^-\hat{\sigma}_{i'}^+ + \text{h.c.})
    \\
    + &\frac{\zeta_i}{4} (\mathds{1}_i\mathds{1}_{i'}-\mathds{1}_i\sigma^z_{i'}-\sigma^z_i\mathds{1}_{i'}+\sigma^z_i\sigma^z_{i'})
    \big].
    \end{split}
\end{equation}
Here, $\sigma^{\pm}$ are the raising and lowering operators, $\sigma^z$ denotes the Pauli Z operator, $\mathds{1}$ denotes the identity and $i'\equiv(i \text{ mod 6})+1$ indicates the index which succeeds $i$.
Due to the anharmonicity of the transmon, hybridisation between states with two excitations results in unwanted ZZ interactions~\cite{Strauch2003, Sung2021}, whose strength $\zeta_i$ depends on the biasing $\phi_\text{dc}$ of the couplers.
The effective coupling strengths $J_i$ from modulating the $k$-th harmonic between adjacent qubits $i$ and $i'$ can be tuned by changing the amplitude $A_i^k$ of the respective parametric drive, resulting in~\cite{Mckay2016, Huber2024} 
\begin{equation}
    \label{eq:g_param}
    J_i \approx \left. \frac{\partial^k \omega_{\text{c}_i}}{{\partial \phi}^k}\right|_{\phi_\text{dc}} \frac{g_{ii} g_{i'i}}{\Delta_i^2} \frac{A_i^k}{2}.
\end{equation}
The ranges of achievable experimental couplings for our devices, shown in Fig.~\ref{fig:device}\textbf{d}, are limited by the onset of higher-order error processes at large modulation amplitudes~\cite{Petrescu2023}.
Without parametric drive, the effective couplings between adjacent qubits are largely suppressed, with ratios $g/\Delta<0.02$ for all pairs, except for qubits $\text{q}_3$ and $\text{q}_4$ which partially hybridise ($g/\Delta=0.16$) due to their small frequency difference.
Therefore, utilising parametric drives enables us to control the active couplings on the device and their relative strengths.

\subsection*{Perfect State Transfer Protocol}
In a chain of $N$ coupled qubits 
\begin{equation}
    \label{eq:H_chain}
    \hat{H}_{\text{chain}} / \hbar = \sum_{n=1}^{N-1}J_n(\hat{\sigma}_n^-\hat{\sigma}_{n+1}^+ + \text{h.c.}),
\end{equation}
a PST is realised by setting the coupling strengths as
\begin{equation}
    \label{eq:PST_couplings}
J_n=\frac{\pi}{2\tau}\sqrt{n(N-n)},
\end{equation}
resulting in a state transfer from any qubit and its mirror-symmetric qubit in the transfer time $\tau$. 
Notably, PST provides the time-optimal solution to transfer states between the ends of the chain~\cite{Yung2006, Kay2022}.

We implement PST on a chain of $N=6$ qubits with a transfer time of $\tau=\SI{640}{\ns}$ by setting the coupling strengths according to the PST formula in Eq.~\eqref{eq:PST_couplings}, as shown in Fig.~\ref{fig:device}\textbf{d}.
The coupling strengths are initially calibrated by individually driving each coupler and sweeping the respective drive amplitude.
Then, all drives are applied simultaneously and their frequencies and amplitudes are further optimised using a closed-loop routine~\cite{Werninghaus2021, Glaser2024} to compensate for shifts caused by neighbouring drives (see Methods).

\begin{figure*}[t]
\includegraphics{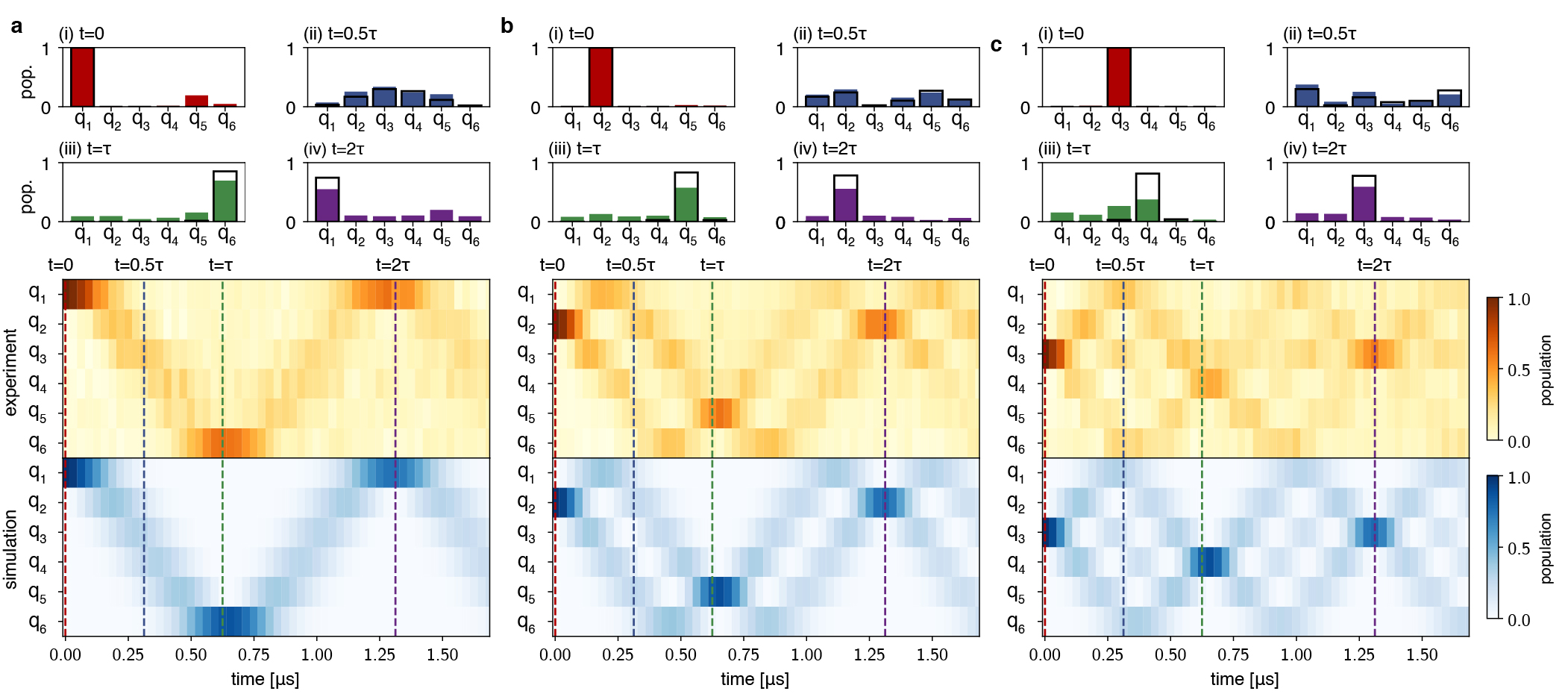}
\caption{
    \label{fig:PST}
    \textbf{Perfect State Transfer protocol.}
    Single-excitation dynamics are shown for excitations starting on an outer~\textbf{a}, intermediate \textbf{b}~and centre \textbf{c}~qubit for a chain of six qubits.
    Experimentally measured populations are visualised in a contour plot (middle) and compared to simulations (bottom), which include the effect of relaxation on the qubits.
    The solid-bar plots (i-iv) show the measured population in all qubits at times $t = 0, 0.5\tau, \tau, 2\tau$ for $\tau=\SI{640}{ns}$.
    Black wireframes indicate the expected populations from simulation.
}
\end{figure*}

We vary the duration of the applied drives and observe transfer dynamics for different initial locations by measuring the excited state population on all qubits.
When the excitation is prepared on one of the outer qubits of the chain, it spreads out to the neighbouring qubits forming a single wave-packet structure, as shown in Fig.~\ref{fig:PST}\textbf{a} when qubit $\text{q}_1$ is initially excited.
At the transfer time $\tau=\SI{640}{\nano\second}$, the excitation refocuses on qubit $\text{q}_6$, located at the other end of the chain. 
The process repeats every integer multiple of the transfer time, with the excitation alternating between the two ends of the chain.
Relevant stages of the transfer dynamics are highlighted in subplots (i-iv).
If instead the excitation is prepared on one of the intermediate qubits, as shown in Fig.~\ref{fig:PST}\textbf{b} for $\text{q}_2$, the excitation first spreads out in two separate wave-packets travelling in opposite directions.
Both components then eventually reflect off the closed boundaries and refocus at the mirror-symmetric qubit, $\text{q}_5$, after the transfer time $\tau$.
The process repeats itself until decoherence effects of the qubits become dominant.
Finally, when preparing the excitation on one of the centre qubits, as shown in Fig.~\ref{fig:PST}\textbf{c} for $\text{q}_3$, the dynamics exhibit multiple splittings, yet the excitation refocuses at integer multiples of $\tau$, as in the other cases. 

We simulate the perfect dynamics of PST by evolving the chain Hamiltonian in Eq.~\eqref{eq:H_chain} in the single-excitation manifold with the ideal coupling strengths $J_n$ from Eq.~\eqref{eq:PST_couplings}.
Relaxation effects are included with the addition of non-Hermitian diagonal terms $-i\pi\hbar\Gamma_1^{n}$, where $\Gamma_1^{n}=1/T_1^{n}$ is the measured decay rate of qubit $\text{q}_n$~\cite{Magnard2018}.
The simulation (blue contour plots in Fig.~\ref{fig:PST}) matches well the observed dynamics, suggesting that errors in the transfer are dominated by decoherence.
Nonetheless, the excitations partially disperse throughout the chain over time, as can be observed when comparing the qubit populations at times $t=0,0.5\tau,\tau,2\tau$ in subplots (i-iv) to simulation results, shown as black wireframes.
While dephasing and flux noise are the main causes of dispersion during the transfer, we attribute this effect also to the hybridisation between qubits $\text{q}_3$ and $\text{q}_4$.

Equivalent results to the ones shown here are obtained when repeating the same experiment for all initial states and chain lengths varying from three to six qubits (see~\ref{sec:all_PST}).
However, notable is the case of the excitation starting on the centre qubit which for odd length chains is the mirror of itself.
In this scenario, the excitation fans out into multiple branches and refocuses back at the centre qubit after every transfer time $\tau$.

\begin{figure}[t!]
    \centering
    \includegraphics{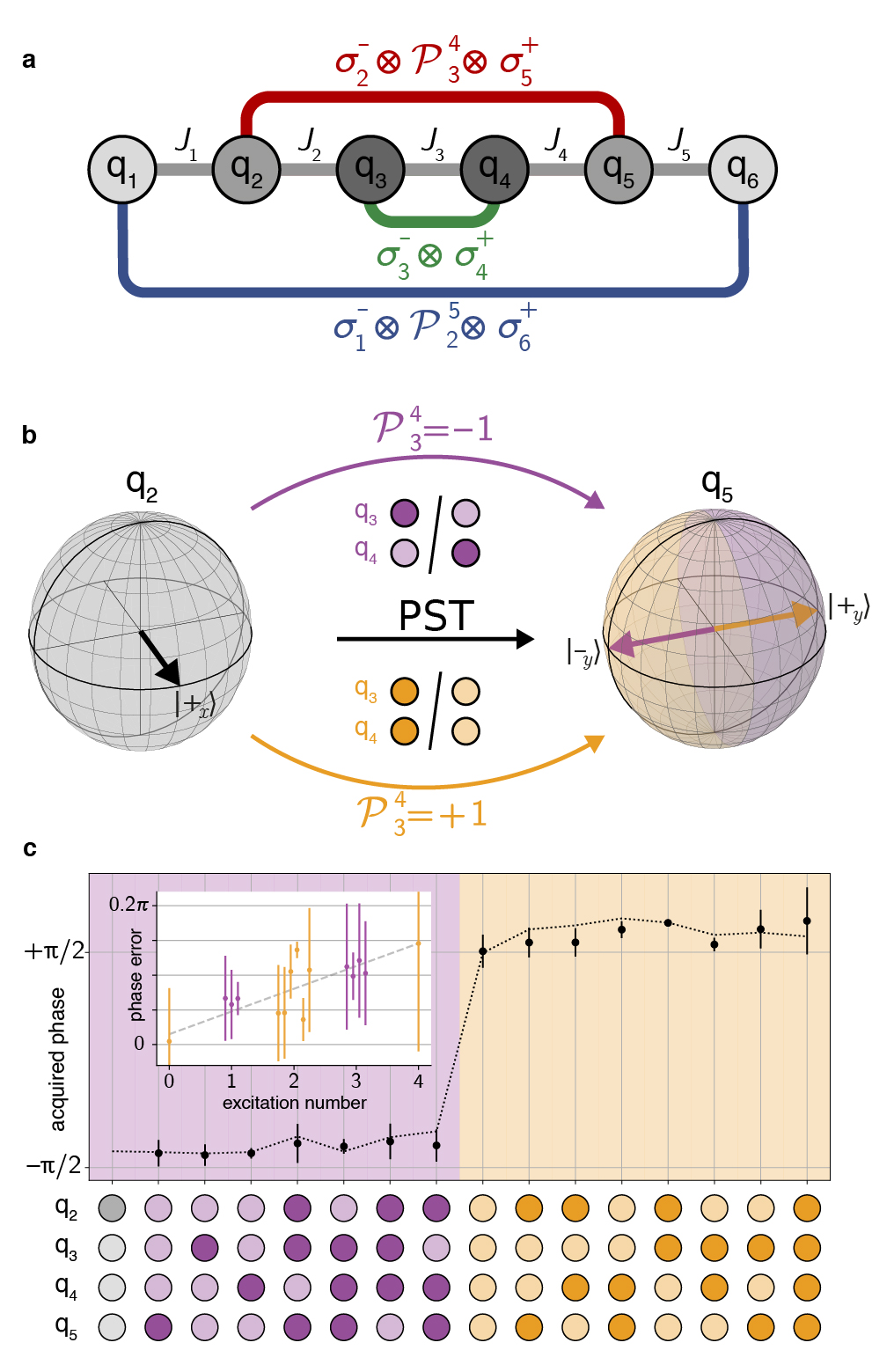}
    \caption{
    \textbf{Parity-dependence of the PST protocol with multiple excitations on a six-qubit chain.}
    \textbf{a}~Schematic of the qubit chain with effective PST couplings.
    Qubits (circles) are connected in a chain by nearest neighbour couplings (grey lines).
    The effective multi-qubit interactions contain hopping terms $\sigma^-\sigma^+$ between the outer (blue), intermediate (red) and centre qubits (green), and include additionally the parity operator applied to all qubits in between.
    \textbf{b}~A superposition state $\ket{+_x}$ prepared on qubit $\text{q}_2$ (left) acquires a phase of $\pm \pi/2$ when transferred to $\text{q}_5$ (right). The sign of the phase is controlled by the number parity $\hat{\mathcal{P}}_{3}^{4}$ of the excitations on qubits $\text{q}_3$ and $\text{q}_4$ (centre).
    \textbf{c}~Acquired phase in the transfer from qubit $\text{q}_1$ to $\text{q}_6$ for all configurations of the inner qubits $\text{q}_2$, $\text{q}_3$, $\text{q}_4$ and $\text{q}_5$. 
    The phase is given by the $x\text{-}y$ angle of the state of $\text{q}_6$ after the transfer, as measured by quantum state tomography.
    Values and error bars (standard deviation) are obtained from the phases measured from four different initial superposition states of $\text{q}_1$, $\ket{\pm_x}$ and $\ket{\pm_y}$.
    The background and label colours highlight the parity (purple - odd, orange - even) of the inner qubits and the excitation of each qubit (light - zero, dark - one).
    Dotted line shows simulation results when taking unwanted ZZ couplings into account.
    Inset shows the distance from ideal phase as function of number of excitations.
    \vspace{-3mm}
    }
    \label{fig:Parity}
\end{figure}

\subsection*{Parity-dependence of Perfect State Transfer}\label{sec:parity-dependence}
The PST protocol produces effective non-local interactions which efficiently transfer single excitations between two distant qubits.
However, in the presence of excitations in multiple qubits along the chain, it produces additional interaction terms, therefore effectively coupling all the qubits in the chain. 
Indeed, the chain Hamiltonian from Eq.~\eqref{eq:H_chain}, shown as grey lines in Fig.~\ref{fig:Parity}\textbf{a}, produces stroboscopically equivalent dynamics to the effective non-local parity-dependent Hamiltonian
\begin{equation}
    \label{eq:H_PST}
    \hat{H}_{\text{PST}}/\hbar = \frac{\pi}{2\tau} \sum_{n=1}^{\lfloor\frac{N}{2}\rfloor}\bigg( \bigotimes_{k=n+1}^{\tilde{n}-1}\hat{\sigma}_k^z\bigg)(\hat{\sigma}_n^-\hat{\sigma}_{\tilde{n}}^+ + \text{h.c.}),
\end{equation}
shown in Fig.~\ref{fig:Parity}\textbf{a} as coloured lines~\cite{Naegele2022}.
Since the dynamics of the two Hamiltonians are equivalent at multiples of the transfer time $\tau$, we can understand the operation generated by the PST protocol through the effective Hamiltonian in Eq.~\eqref{eq:H_PST}.
Here, the transversal coupling terms $\sigma^-\sigma^+$ in the sum implement the transfers between the mirror-symmetric qubits at positions $n$ and $\tilde{n}=N+1-n$.
At the same time, the parity operators $\hat{\mathcal{P}}_{n+1}^{\tilde{n}-1}=\bigotimes_{k=n+1}^{\tilde{n}-1}\hat{\sigma}^z_k \in \{-1, 1\}$ modify the sign of the coupling depending on the number parity of excitations between qubits $\text{q}_n$ and $\text{q}_{\tilde{n}}$.
Therefore, each mirror-symmetric transfer produces a parity-dependent phase of $\pm \pi/2$, resulting in the effective multi-qubit operation
\begin{equation}
    \label{eq:iswap}
    \hat{\mathcal{P}}\text{-}i\text{SWAP}_n=
    \begin{bmatrix}
        1 & 0 & 0 & 0\\
        0 & 0 & e^{i\hat{\mathcal{P}}_{n+1}^{\tilde{n}-1}\pi/2} & 0\\
        0 & e^{i\hat{\mathcal{P}}_{n+1}^{\tilde{n}-1}\pi/2}& 0 & 0\\
        0 & 0 & 0 & 1\\
    \end{bmatrix}.
\end{equation}
The full PST operation is $U_\text{PST}=\bigotimes_{n=1}^{\lfloor\frac{N}{2}\rfloor} \hat{\mathcal{P}}\text{-}i\text{SWAP}_n$.
To intuitively understand the origin of this parity-dependence, we can think of the system of qubits as fermions in a lattice by using a Jordan-Wigner transform~\cite{Jordan1928, Ortiz2001}.
When fermions swap through each other they acquire a phase of $\pi$ changing the phase of the transferred state from $\pi/2$ to $-\pi/2$ and vice versa.
Since the Jordan-Wigner transform leaves the chain Hamiltonian from Eq.~\eqref{eq:H_chain} unchanged, the fermion and qubit dynamic are equivalent, with the qubit experiencing the same phase changes.

The parity-dependence can be observed explicitly by monitoring the evolution of the superposition state $\ket{+_x}_n\equiv(\ket{0}_n+\ket{1}_n)/\sqrt{2}$, prepared on qubit $\text{q}_n$.
When PST is applied, the state is transferred to $\text{q}_{\tilde{n}}$ and acquires a phase conditioned on the number of encountered excitations, i.e. $\ket{+_x}_n\rightarrow\ket{\pm_y}_{\tilde{n}}=(\ket{0}_{\tilde{n}} \pm i\ket{1}_{\tilde{n}})/\sqrt{2}$, where the sign depends on the number parity.
This mechanism is shown on the Bloch sphere in Fig.~\ref{fig:Parity}\textbf{b} for a transfer from qubit $\text{q}_2$ to $\text{q}_5$ on a $N=6$ qubit chain.

In our experiment, we verify the parity-dependent property with the transfer between the ends of the chain.
We prepare qubit $\text{q}_1$ in a superposition state, apply PST and determine the state of qubit $\text{q}_6$ using quantum state tomography~\cite{Chuang1998, Liu2005, Steffen2006}. 
The density matrix is reconstructed from the tomographic data using a maximum-likelihood estimator while imposing physical constraints~\cite{Hradil2004}.
The $x\text{-}y$ angle of the transferred state is then measured for all possible initial computational states in the inner qubits $\text{q}_2$, $\text{q}_3$, $\text{q}_4$ and $\text{q}_5$.
We repeat the process for four different initial superposition states of $\text{q}_1$, $\ket{\pm_x}$ and $\ket{\pm_y}$, and for each determine the phase acquired during the transfer, as shown in Fig.~\ref{fig:Parity}\textbf{c}.
As expected, the phase of the transferred state encodes the parity of the inner qubits, corresponding to $\sim\pi/2$ for an even number of excitations and to $\sim-\pi/2$ when the number parity is odd.
We attribute unwanted ZZ interactions to cause deviations from the ideal $\pm\pi/2$ phase.
This is supported by the fact that the error in the acquire phase increases linearly with the number of excitation in the inner qubits, as shown in the inset of Fig.~\ref{fig:Parity}\textbf{c}.
Furthermore, we simulate the PST operation including ZZ interaction with measured strengths $\zeta_i$ obtaining good agreement with the observed results [dotted line in Fig.~\ref{fig:Parity}\textbf{c}].

\subsection*{Entanglement generation}
The parity-dependent property of the PST makes it a powerful tool for generating entanglement between the qubits in a chain.
Each of the effective transfer processes that occurs during PST yields correlations between the state of the transferred qubits and the state of the qubits between them.
Therefore, by exploiting the correlations produced by these simultaneous transfers, we can generate entanglement over the whole chain~\cite{Clark2005}.
This intuition is formalised by using the theory of graph states, a subset of multi-qubit entangled states which are well studied and allow for a simple graphical representation~\cite{Hein2004, DenNest2004, Hein2006}.
In a graph state, nodes represent qubits initialised in the $\ket{+}$ state and edges indicate pairwise Ising $\sigma^z\sigma^z$ (CZ-like) interactions between the nodes.
In this formalism, a GHZ state of the form $(\ket{0\dots0}+\ket{1\dots1})/\sqrt{2}$ corresponds up to single-qubit operations to a complete graph, i.e. with all-to-all connectivity. 
Therefore, a GHZ state can be achieved by preparing the qubits in a superposition state, applying consecutive CZ gates between all pairs of individual qubits and applying a final layer of single qubit gates.

\begin{figure}[t!]
    \label{fig:GHZ}
    \centering
    \includegraphics{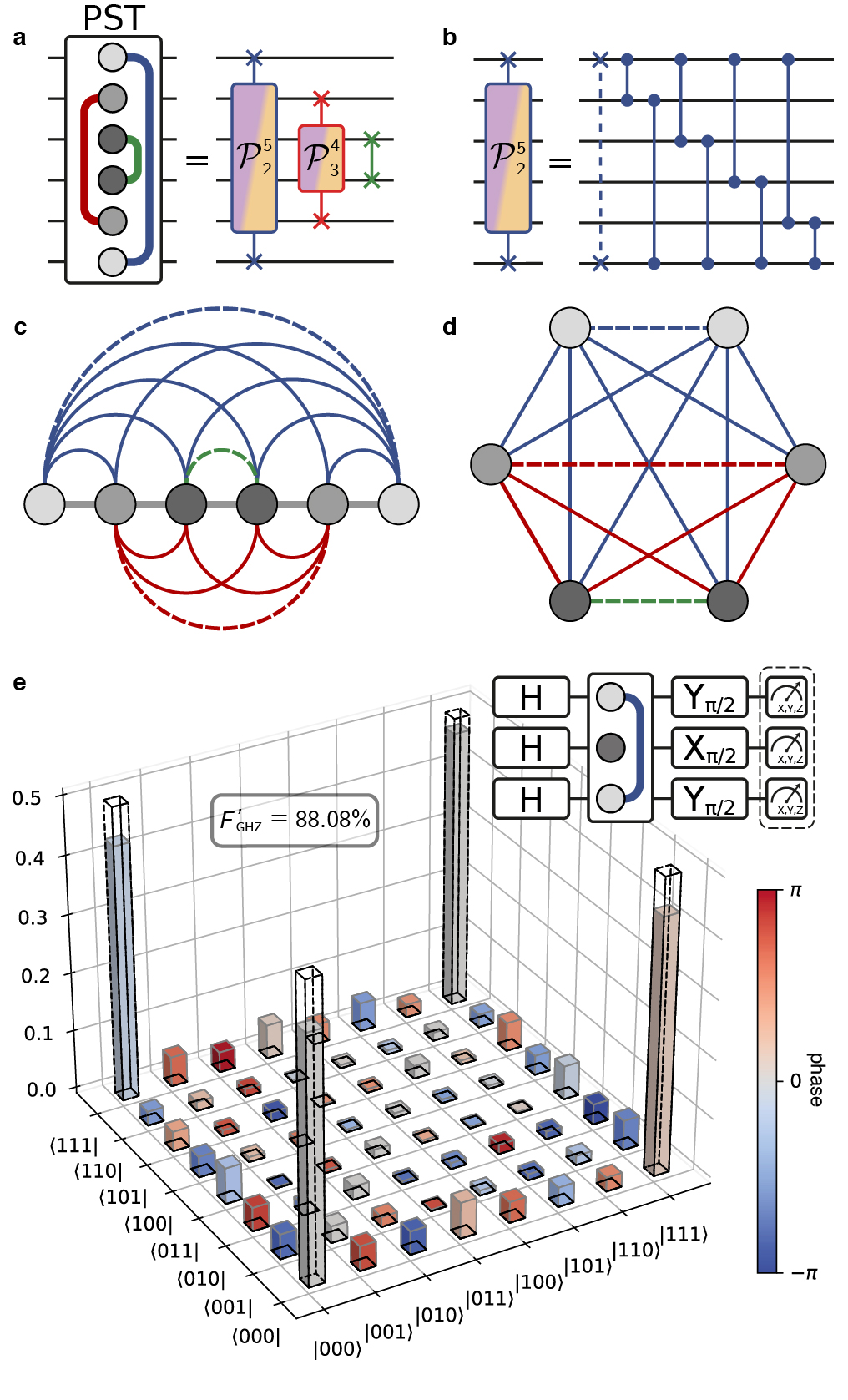}
    \caption{
    \textbf{Multi-qubit entanglement generation with a single PST operation.}
    Circuit decomposition of the six-qubit PST into iSWAP operations with parity-controlled phases \textbf{a}~which each decompose further into an iSWAP and multiple CZ gates~\textbf{b}. 
    \textbf{c}~Graph representation of the chain with edges corresponding to all effective two-qubit interactions generated by a single PST.
    Dashed edges denote iSWAP interactions, whereas continuous lines indicate CZ interactions.
    Colours identify the specific transfer from which each two-qubit interaction originates.
    \textbf{d}~Rearrangement of the chain graph which emphasises the all-to-all connectivity corresponding to a GHZ state.
    \textbf{e}~Reconstructed density matrix of the experimentally realised GHZ state $\rho_{\text{exp}}$ (solid bars), with ideal values plotted as black wireframes $\rho_\text{ideal}=\ket{\psi_\text{GHZ}}\bra{\psi_\text{GHZ}}$. 
    Allowing for an arbitrary Z rotation of the final state yields $F'_\text{GHZ}=\SI{88.08}{\%}$.
    Inset shows the circuit diagram for the GHZ state creation on a chain of size $N=3$.
    }
\end{figure}

To illustrate how the PST operation maps to the graph state formalism, we decompose it into two-body operations.
As derived in Eq.~\eqref{eq:iswap} and shown in Fig.~\ref{fig:GHZ}\textbf{a} for six qubits, the PST can be described by a product of mirror-symmetric transfers, each implementing a parity-dependent iSWAP operation.
Each mirror-symmetric transfer can then be decomposed into a single iSWAP gate and a series of CZ gates implementing the parity-dependence, as shown in Fig.~\ref{fig:GHZ}\textbf{b}. 
In the graph state representation, each of these two-qubit interactions contributes individual edges between the qubits in the chain, resulting in the all-to-all connected graph shown in Fig.~\ref{fig:GHZ}\textbf{c}-\textbf{d}.
As this remains valid for different qubit numbers, a single PST operation can be utilised to directly generate a GHZ state for any chain length.
Note that the iSWAP gates also contribute edges to the graph, since these are equivalent to a CZ gate followed by a SWAP operation and single-qubit Z gates:
Given that all qubits are initialised in the $\ket{+_x}$ state, the SWAP operations do not alter the state of the chain, leaving only the Z gates, which commute with the CZ operations and are incorporated in the final layer of single-qubit gates (see~\ref{sec:ghz_circuit}).

We demonstrate this process for a chain of three qubits $\text{q}_5$, $\text{q}_6$ and $\text{q}_1$, using the gate sequence shown in the inset of Fig.~\ref{fig:GHZ}\textbf{e}, where the PST operation lasts $\tau=\SI{390}{\nano\second}$.
These qubits were chosen to mitigate the adverse impact of residual ZZ interactions, which would affect the PST operation as well as single-qubit gate fidelities.
All qubits are prepared in an equal superposition state $\ket{+_x}$ by applying Hadamard gates.
Applying a single PST operation entangles the state of the qubits by imparting parity-dependent phases to each of the computational states.
Finally, a layer of single-qubit gates maps the fully-entangled graph state onto the GHZ state $\ket{\psi_\text{GHZ}} = (\ket{000} + \ket{111})/\sqrt{2}$.
The obtained state $\rho_\text{exp}$ is determined using quantum state tomography and produces a $F_\text{GHZ}=\text{Tr}\qty(\rho_{\text{exp}}^\dagger \rho_{\text{ideal}})=\SI{84.97}{\%}$ fidelity overlap with the targeted GHZ state, as shown in Fig.~\ref{fig:GHZ}\textbf{e}.
The PST dynamics also result in an additional phase of $e^{i0.378}$ on the $\ket{111}$ state. 
Since the GHZ state is a superposition of only two computational states, this phase error is equivalent to applying a single-qubit Z rotation.
When allowing for this additional phase freedom in the GHZ, we obtain a fidelity of $F'_\text{GHZ}=\SI{88.08}{\%}$.
The obtained fidelity is close to the simulated decoherence limit of \SI{89.11}{\%}.

\section*{Discussion}
We have experimentally implemented a Perfect State Transfer (PST) protocol on chains of up to six qubits, demonstrating simultaneous control of parametric couplings of multiple neighbouring qubit pairs.
This state transfer occurs not only between the qubits at the end of the chain, but also for all mirror-symmetric pairs. 
Furthermore, in the presence of multiple excitations along the chain, our experiments exhibit parity-dependent properties where the phase of a transferred state depends on the number parity of all excitations within the chain, in good agreement with the theoretical prediction. 
Harnessing the parity-dependence, we explicitly demonstrate the generation of a GHZ state in a chain of three qubits using a single PST operation with a fidelity of \SI{88.08}{\%}.
By mapping the entanglement generation protocol to the theory of graph state, the method can be generalised to larger qubit numbers.

The two main sources of errors in our implementation of PST are decoherence during the gate and residual ZZ couplings.
The effect of decoherence in our system can be overcome by increasing the strength of the static capacitive couplings $g_{ij}$ to achieve faster transfers, as well as by increasing the $E_J/E_C$ ratio to suppress dephasing due to charge noise. 
Precise targeting of qubit frequencies would allow to operate the qubits in the straddling regime $(|\omega_{\text{q}_i}-\omega_{\text{q}_j}|<-\alpha_{{\text{q}_i/\text{q}_j}})$ where the ZZ coupling can be fully suppressed~\cite{Sete2021, Sung2021}.
Alternatively, additional drives can be used to control and cancel out these unwanted couplings~\cite{Ni2022, Ganzhorn2019, Noguchi2020, Wei2022}.

By harnessing simultaneous interactions with no additional all-to-one resources, PST provides an efficient way to implement effective operations between distant qubits on the same or on different chips and create many-qubit entanglement.
Compared to its decomposition into single- and two-qubit gates, PST results in a two-fold reduction in total gate time~\cite{Naegele2022} and exhibits increased robustness to coherent and incoherent errors (see~\ref{sec:Robustness} for details), presenting a useful tool in the pursuit to create large-scale quantum computers.
For example, PST could be used for applications in parity-check codes.
Here, the transfer and measurement of a superposition state between the outer qubits in a chain, as implemented in this work, realises a direct parity measurement of the other $N-2$ qubits.
Furthermore, realising PST on overlapping chains opens up the possibility to generate different classes of graph state~\cite{Angelakis2008} and enable quantum routing~\cite{Dutta2023}, with potential applications in quantum communication~\cite{Azuma2015, Borregaard2020} and quantum sensing~\cite{Shettell2020}. 
Besides purely unitary operations, we note that GHZ states can also be generated in constant depth via measurement-based circuits~\cite{Riste2013, Buhrman2024, Foss2023, Koh2024, Alam2024, Baumer2024}.

In addition to the discussed applications, the PST protocol can be extended and modified in a number of ways.
By introducing detunings between the qubits, the method can be generalised to implement Fractional State Transfer~\cite{Dai2010, Banchi2015, Genest2016, Lemay2016, Chan2019, Naegele2022}, where excitations only partially transfer between mirror-symmetric qubits.
The fractional transfer operation exhibits the same parity-dependent property of PST (see experimental results in~\ref{sec:FST}), which allows for performing small evolution steps of the effective parity-dependent interactions, Eq.~\eqref{eq:H_PST}, enabling simulation of fermionic systems~\cite{Arute2020} and gauge field theories~\cite{Mildenberger2025}.
Chain Hamiltonians can also be tailored to implement more general types of transfers, e.g. from one qubit to many, thus allowing for the generation of larger classes of entangled states~\cite{Kay2017, Kay2017_tailoring}.
Moreover, extending the couplings to allow for time-dependent control provides further speed-ups~\cite{Caneva2009} and enables the exploration of a large family of multi-qubit operations~\cite{Naegele2022_thesis}.
Finally, PST can be generalised to operate in networks with higher connectivity: 
analytical solutions for coupling strengths resulting in PST have been found for hyper-cubes and other cube-like graphs~\cite{Christandl2005, Facer2006, Bernasconi2008}, triangular lattices~\cite{Miki2012, Post2015, Miki2022}, multi-layer hexagonal lattices~\cite{Karimipour2012}, as well as any network which can be described as intersections of one-dimensional chains~\cite{Pemberton-Ross2011, Kay2010}.
The latter results provide a direct implementation in higher-dimensional grid lattices, 
as shown in~\ref{sec:PST2D}.

\section*{Methods}
\subsection*{Calibration of parametric drives}
\label{sec:PST_calib}
To control the strength of the simultaneous parametric couplings, we perform multiple rounds of calibration.
First, we characterise the effective coupling strengths $J_i$ for each pair of neighbouring qubits as a function of the parametric drive amplitude $A_i$ while all other drives $(A_{j \neq i})$ are set to zero. 
For each amplitude, we measure the population of the two qubits involved while varying the drive duration and frequency. 
The resulting Chevron patterns are then fitted to Rabi oscillations in order to estimate the effective coupling strengths $J_i(A_i)$ and the resonant transition frequencies.
The obtained values deviate slightly from Eq.~\eqref{eq:g_param} due to frequency collisions and higher-order corrections.
In particular, the transition frequencies are shifted from the expected qubit-qubit detunings $\Delta_i$ due to drive-induced Stark shifts.
Given the required PST coupling strengths from Eq.~\eqref{eq:PST_couplings}, we are then able to set the drive amplitude and frequency accordingly.
For each pair, we repeat the same procedure while also driving the two neighbouring couplers off-resonantly.
This allows us to correct up to first order the qubit frequency shifts caused by neighbouring simultaneous drives.

\begin{figure}[t!]
   \centering
    \includegraphics{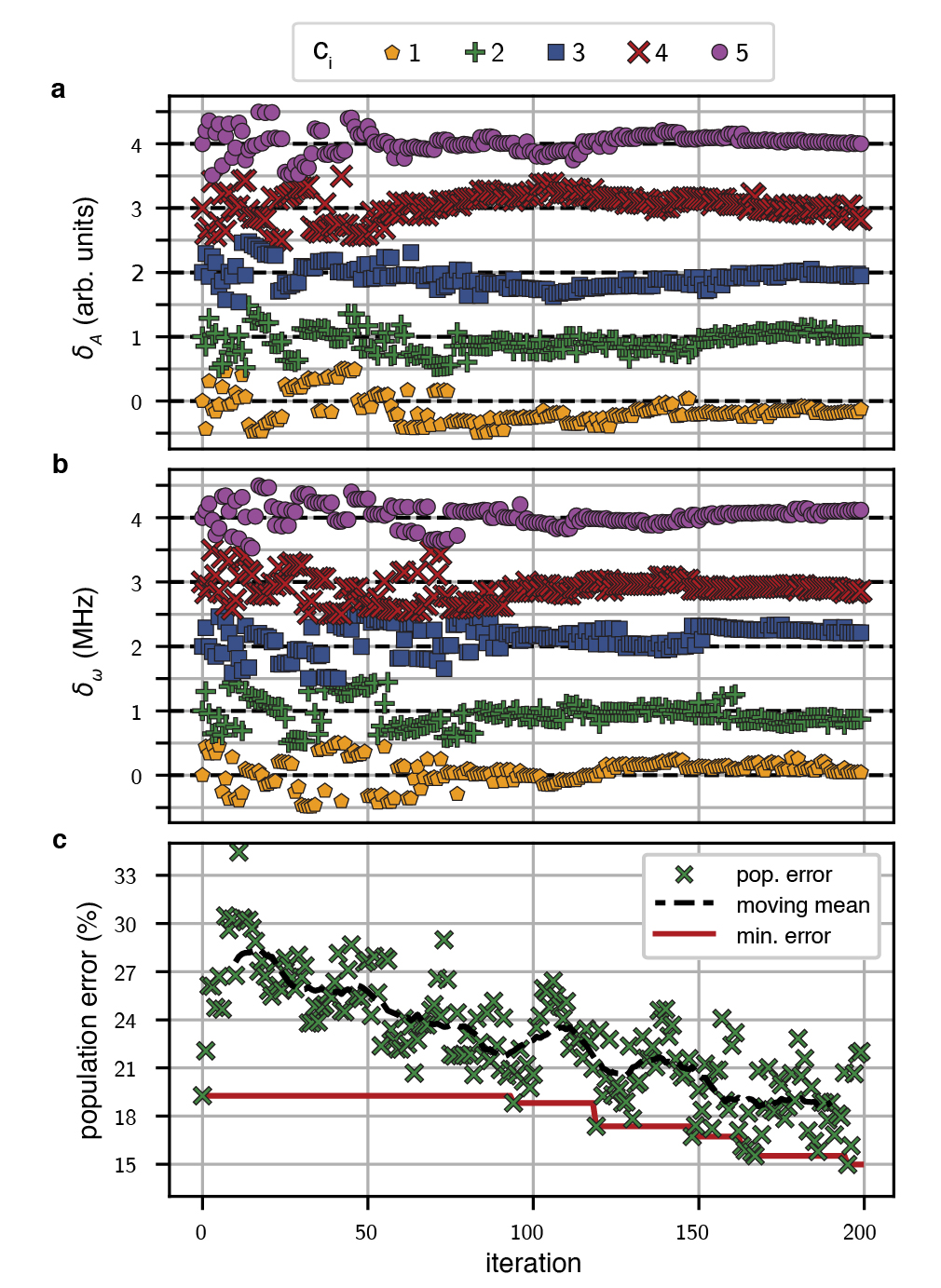}
    \caption{
    \textbf{Optimisation of simultaneous drives for PST on the six-qubit chain.}
    Parameter changes as a function of optimiser iteration are shown for drive amplitudes \textbf{a}~and drive frequencies~\textbf{b}.
    Results for the five parametric drives are offset for visualisation, with dashed black lines showing their initial value.
    For reference, the average drive amplitude is 0.21 (arb. units) and the average drive frequency is \SI{183}{\MHz}.
    \textbf{c}~Population errors averaged over all qubits over up to five consecutive transfers.
    Black dashed line and red line indicate the moving mean and minimum of the error respectively.
    \vspace{-3mm}
    }
    \label{fig:app_optimization}
\end{figure}

As a final step, we apply all drives needed for PST simultaneously and fine-tune all amplitudes and frequencies using an optimiser enabled with experiment feedback.
The experiment consists of transferring a single excitation from any initial qubit, with populations in all qubits measured at intervals up to five times the transfer time $\tau$. 
The resulting average population error, calculated from an ideal transfer, is then fed to a Tree-Structured Parzen Estimator optimiser provided in the Optuna Python library~\cite{Akiba2019}.
Convergence of the closed-loop optimiser is shown in Fig.~\ref{fig:app_optimization} for transferring an excitation initially prepared on qubit $\text{q}_1$ through a chain of six qubits.
Convergence of the closed-loop optimiser results in drive parameters accounting for all cross-dependencies and thus reducing the transfer error, as shown in Fig.~\ref{fig:app_optimization} for a chain of $N=6$ qubits.

\section*{Data Availability}
All relevant data supporting the main conclusions and figures of the document are available upon request. 


\bibliography{ncomms_final_arxiv}


\section*{Acknowledgments}
We thank Ignacio Cirac, Peter Rabl and Xavier Coiteux-Roy for insightful discussions and helpful comments.
This work received financial support by the European Union’s Horizon 2020 research,  
the innovation program `MOlecular Quantum Simulations' (MOQS; Nr.~955479),
the EU MSCA Cofund `International, interdisciplinary and intersectoral doctoral program in Quantum Science and Technologies' (QUSTEC; Nr.~847471),
the BMBF programs `German Quantum Computer based on Superconducting Qubits' (GeQCoS; Nr.~13N15680) and MUNIQC-SC (Nr.~13N16188),
the German Research Foundation project `Multi-qubit gates for the efficient exploration of Hilbert space with superconducting qubit systems' (Nr.~445948657) and the excellence initiative `Munich Center for Quantum Science and Technology' (MCQST; Nr.~390814868) as well as the Munich Quantum Valley, which is supported by the Bavarian state government with funds from the Hightech Agenda Bayern Plus. 
F.P. has received funding from the BMW group and C. Schw. has received funding from LMUexcellent.

\section*{Author contributions}
F.R. and J.R. designed and carried out the experiments and analysed the data.
F.R., J.R. and K.L. performed numerical simulation.
I.T. and G.H. designed the chip.
L.K., N.B., D.B., L.R. and L.S. fabricated the chip.
J.S. designed and procured the cryoperm shields, the PCB and the chip housing.
N.G., C.Schw., M.W., M.S., F.P., F.R. and J.R. developed the measurement software framework.
F.H, A.M., G.K., C.Schn., F.W., C.Schw., I.T., M.S., L.K., F.R. and J.R. built and maintained the experimental setup.
S.F. and M.W. supervised the project.

\section*{Competing Interests}
The authors declare no competing interests.

\vfill

\pagebreak
\onecolumn
\section*{\fontsize{14.5bp}{14bp}\bfseries\titraggedcenter Parity-dependent state transfer for direct entanglement generation \\-- Supplementary Information --}
\vspace{3mm}
\newcommand\myminus{\kern1.5pt\rule[3pt]{4pt}{0.4pt}\kern1pt}
\newcommand\myminustable{\kern1pt\rule[2.2pt]{3pt}{0.3pt}\kern0.5pt}
\sisetup{separate-uncertainty=true,range-units=single,range-phrase=\myminustable}
\renewcommand{\figurename}{Supplementary Fig.}
\renewcommand{\theHfigure}{Supplementary Fig.}
\setcounter{figure}{0}
\renewcommand{\tablename}{Supplementary Table}
\renewcommand{\thesubsection}{Supplementary Note \arabic{subsection}}
\renewcommand{\theHsubsection}{\arabic{subsection}}

\begin{figure}[b]
    \centering
    \includegraphics{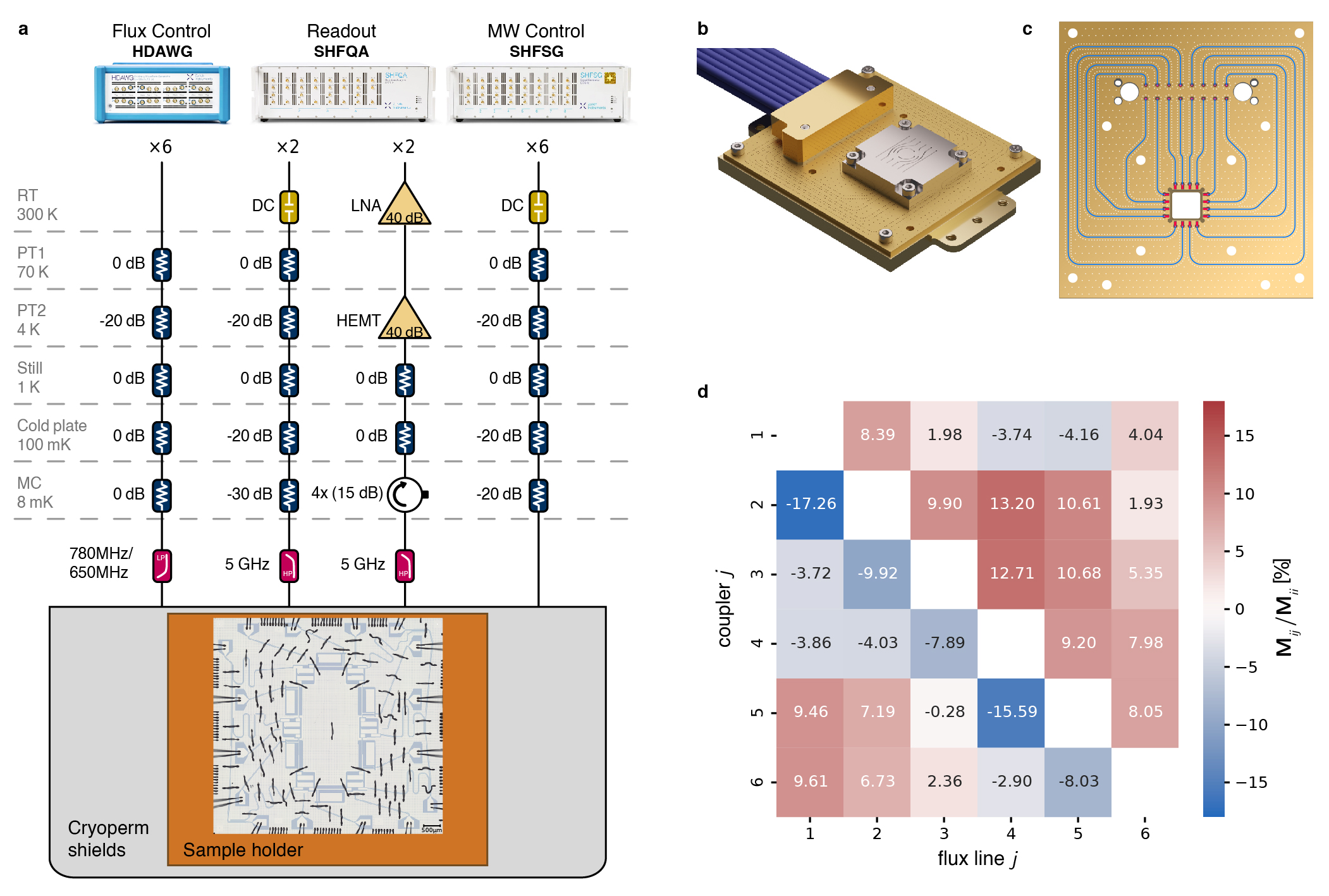}
    \caption{
    \label{fig:fridge_setup}
    $\mybar$ \textbf{Experimental setup and DC flux crosstalk.}
    \textbf{a}~Wiring and control electronics of the experiment. 
    Schematic of the qubit housing showing the PCB with Ardent connector and copper sample holder~\textbf{b} and the routing of the PCB~\textbf{c}.
    See main text for details.
    \textbf{d}~Flux crosstalk matrix: entries correspond to the ratio $\textbf{M}_{ij}/\textbf{M}_{ii}$, given in percentage, where $\textbf{M}_{ij}$ is the required change in voltage applied to line $j$ in order to induce a change of one flux quantum in the flux threading the SQUID loop of coupler $i$.
    }
\end{figure}

\subsection{-- Experimental setup}
\label{sec:exp_setup}
A schematic of the experimental setup is shown in Supplementary Fig.~\ref{fig:fridge_setup}\textbf{a}.
The qubit MW control pulses are generated using a Zurich Instruments (ZI) ``Super High Frequency Quantum Analyzer" (SHFSG).
The MW drive lines are attenuated by \SI{-60}{\dB} distributed over the different temperature stages and routed to the individual qubit drive lines on the device.
Flux biasing and parametric drives are applied simultaneously using a ZI ``High Density Arbitrary Waveform Generator" (HDAWG).
The flux lines are attenuated by \SI{-20}{\dB}, filtered using a \SI{780}{\MHz} or \SI{650}{\MHz} low-pass filter (Mini-Circuits VLFX-780+ or VLFX-650+) and routed to the individual coupler flux lines on the device.
The measurement pulses are generated using a ZI ``Super High Frequency Quantum Analyzer" (SHFQA), allowing us to multiplex different readout tones on the same signal.
The measurement signal lines are attenuated by \SI{-70}{\dB} distributed over the different temperature stages, filtered using a \SI{5}{\GHz} high-pass filter (Mini-Circuits VHF5050+) and routed to the two feedlines on the device.
The output signals from the feedlines are filtered using a \SI{5}{\GHz} high-pass filter, routed through four \qtyrange[range-phrase=\myminus]{4}{12}{\GHz} isolators and \SI{0}{\dB} attenuators for thermal connection at each stage, then amplified by a \qtyrange[range-phrase=\myminus]{4}{8}{\GHz} \SI{40}{\dB} HEMT cryogenic low-noise amplifier (LNF-LNC4\_8F) on the \SI{4}{\kelvin} stage and a \SI{40}{\dB} low-noise room temperature amplifier (BZ-04000800-081045-152020), before being routed back to the SHFQA for analysis.
The chip is housed in a 16-port package with a copper cavity. Signals are launched to a multi-layer Isola Astra PCB with a 16-port Ardent connector, shown in Supplementary Fig.~\ref{fig:fridge_setup}\textbf{b}-\textbf{c}, and routed from the PCB traces to the chip with wire bonds.
The properties of qubits, resonators and couplers are given in Table~\ref{tab:device}. Qubit coherence times are measured with all tunable couplers positioned at their respective upper sweet spots. Notably, qubits $\text{q}_2$, $\text{q}_5$ and $\text{q}_6$ are affected by charge dispersion due to lower $E_J/E_C$ ratios ($\lesssim 36$), which lead to beating patterns in Ramsey experiments. From these, we extract the decay rate $\Gamma$ and the charge dispersion, resulting in beating frequencies $\Delta$ in the range of $\SIrange{70}{300}{\kHz}$, and determine the effective $T_2^*$ as the time when the envelope $e^{-\Gamma t}\cos(\Delta t)$ first reaches the $1/e$ threshold.

The system exhibits large flux crosstalk due to undesired inductive coupling of SQUID loops to distant flux lines.
We characterise the mutual inductance between all flux lines and couplers by performing spectroscopy measurements on the resonators and qubits, while varying the DC voltage bias on each line~\cite{Abrams2019, Dai2021, Barrett2023}.
The flux experienced by each coupler is given by
\begin{equation}
    \label{eq:app_crosstalk}
    \bar{\Phi} = \textbf{M} \bar{V} + \bar{\Phi}_\text{off}
\end{equation}
where $\bar{V}_j$ is the voltage bias applied to flux line $j$ and $\bar{\Phi}_i$ is the flux experienced by the coupler $i$. 
The matrix element $M_{ij}$ characterises the change in voltage required on line $j$ to vary the flux experienced by coupler $i$ by one flux quantum.
The vector $\bar{\Phi}_\text{off}$ accounts for additional flux sources present in the environment.
For a desired flux operation point of the couplers, the required DC bias $\bar{V}$ is calculated by simply inverting Eq.~\eqref{eq:app_crosstalk}. 
The flux crosstalk in the device is determined by normalising the values $M_{ij}$ by $M_{ii}$, as shown in Supplementary Fig.~\ref{fig:fridge_setup}\textbf{d}.

\begin{table}[t]
\small
\setlength\tabcolsep{2pt}
\begin{tabular}{| c || c | c | c | c | c | c |}
  \hline
  i & 1 & 2 & 3 & 4 & 5 & 6\\ 
 \hline\hline
Qubit frequency, $\omega_{\text{q}}$  & \SI{4.37}{\GHz}  & \SI{3.93}{\GHz}  & \SI{4.27}{\GHz}  & \SI{4.23}{\GHz}  & \SI{3.83}{\GHz}  & \SI{3.21}{\GHz}  \\ \hline
Qubit anharmonicity, $\alpha_{\text{q}}$  & \SI{-247.3}{\MHz}  & \SI{-249.4}{\MHz}  & \SI{-249.7}{\MHz}  & \SI{-252.1}{\MHz}  & \SI{-250.6}{\MHz}  & \SI{-255.1}{\MHz}  \\ \hline
Resonator frequency, $\omega_{\text{r}}$  & \SI{6.04}{\GHz}  & \SI{5.97}{\GHz}  & \SI{6.04}{\GHz}  & \SI{6.11}{\GHz}  & \SI{6.15}{\GHz}  & \SI{6.17}{\GHz}  \\ \hline
Coupler frequency range, $\omega_{\text{c}}$  & \qtyrange{3.65}{7.17}{\GHz} & \qtyrange{4.92}{7.51}{\GHz}& \qtyrange{3.38}{7.28}{\GHz} & \qtyrange{4.66}{6.75}{\GHz} & \qtyrange{2.57}{4.71}{\GHz} & \qtyrange{3.95}{6.93}{\GHz}  \\ \hline
Relaxation time, $T_1$  & \SI{12.1\pm2.4}{\micro\second} & \SI{53.2\pm7.3}{\micro\second} & \SI{26.2\pm3.4}{\micro\second} & \SI{46.0\pm8.6}{\micro\second} & \SI{63.4\pm9.4}{\micro\second} & \SI{72.0\pm7.0}{\micro\second} \\ \hline
Ramsey decay time, $T_2^*$  & \SI{10.0\pm2.0}{\micro\second} & \SI{8.1\pm 6.5}{\micro\second} & \SI{8.1\pm1.1}{\micro\second}  & \SI{7.0\pm 1.0}{\micro\second}  & \SI{4.3\pm 1.7}{\micro\second}  & \SI{4.1\pm 1.2}{\micro\second} 
\\ 
\hline
Readout fidelity $F_{\text{RO}}$ & \SI{78.3 \pm 0.6}{\%} & \SI{91.3 \pm 0.1}{\%} & \SI{89.4 \pm 1.6}{\%} & \SI{87.4 \pm 0.9}{\%} & \SI{81.8 \pm 1.1}{\%} & \SI{80.0 \pm 0.9}{\%} \\

\hline
single-qubit RB error $\epsilon_{1\text{q}}$ & \SI{0.20 \pm 0.02}{\%} & \SI{0.23 \pm 0.01}{\%} & \SI{0.11 \pm 0.02}{\%} & \SI{0.32 \pm 0.06}{\%}  & \SI{0.20 \pm 0.01}{\%} & \SI{0.053 \pm 0.003}{\%} 
\\
\hline
ZZ coupling $\zeta_i$ & \SI{-108\pm 27}{kHz} & \SI{-362\pm 8}{kHz} & \SI{816\pm 14}{kHz} & \SI{-273\pm 8}{kHz} & \SI{-18\pm 27}{kHz} & \SI{-31\pm 26}{kHz}
 \\ 
\hline
Qubit - Next Coupler, $g_{i,j=i}$ & \SI{62}{MHz}  & \SI{74}{MHz}  & \SI{68}{MHz}  & \SI{60}{MHz}  & \SI{47}{MHz}  & \SI{77}{MHz} 
\\ 
\hline
Qubit - Prev. Coupler, $g_{i,j=i-1}$ & \SI{65}{MHz}  & \SI{59}{MHz}  & \SI{112}{MHz} & \SI{65}{MHz} & \SI{61}{MHz}  & \SI{64}{MHz} 
\\ 
\hline
Qubit - Qubit, $g_{i,i+1}$ & $\mbox{--}$ & \SI{6.0}{MHz} & \SI{8.3}{MHz} & \SI{6.6}{MHz} & \SI{4.8}{MHz} & $\mbox{--}$ \\
 \hline

 \end{tabular}
 \caption{
 $\mybar$ \textbf{System parameters.}
 }
\label{tab:device}
\end{table}

\subsection{-- PST in chains with different length}
\label{sec:all_PST}
We perform PST operations on chains of varying length $N$ by applying the calibration method described in the main text to different qubit subsets.
For each calibrated transfer, dynamics were probed for all possible initial locations by applying the simultaneous drive for varying duration and measuring all qubit populations.
Dynamics for exemplary chains composed of three through six qubits are shown in Supplementary Fig.~\ref{fig:all_PST}. 
Equivalently to the subset of plots showed in the main text, these are compared to simulation of the PST Hamiltonian with ideal coupling strengths and including decay rates as non-Hermitian terms.
Calibrated transfer times $\tau_N$ for the chains here depicted are $\tau_{3} = \SI{216}{\ns}$, $\tau_{4} = \SI{429}{\ns}$, $\tau_{5} = \SI{500}{\ns}$ and $\tau_{6} = \SI{640}{\ns}$.

\begin{figure}[b]
    \centering
    \includegraphics{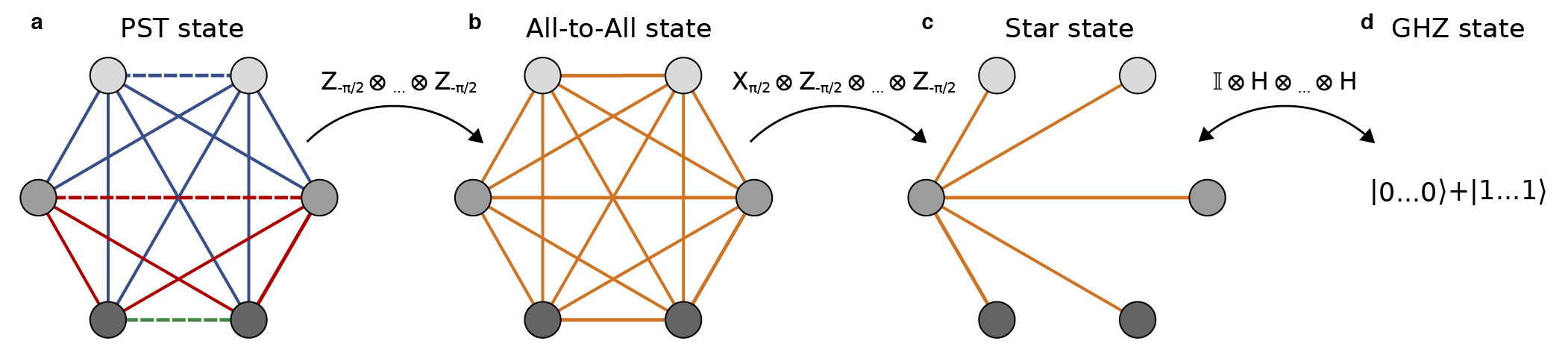}
    \caption{
    \label{fig:mapping}
    $\mybar$ \textbf{Steps of derivation for GHZ circuit with six qubits.}
    By using only single-qubit operations, the entangled state obtained by applying the PST operation to an equal superposition state~\textbf{a} can be mapped to well-known graph states~\textbf{b-c} and ultimately to the GHZ state~\textbf{d}. Labels above each black arrow show the required local unitary for each step.}
\end{figure}

\subsection{-- Circuit for GHZ creation}
\label{sec:ghz_circuit}
As discussed in the main text, PST enables the direct generation of GHZ states.
This is achieved by preparing all qubits in the superposition state, applying PST to realise multiple simultaneous transfers with parity-dependent phases and mapping the resulting state onto the GHZ state via a final layer of single-qubit gates.
We derive the exact gates required in the final layer by using a sequence of local transformations to well-known graph states~\cite{Hein2004}.
Firstly, we map the state generated after the PST operation (Supplementary Fig.~\ref{fig:mapping}\textbf{a}) to the all-to-all connected graph state (Supplementary Fig.~\ref{fig:mapping}\textbf{b}) by applying a $Z_{-\pi/2}$ on all qubits except the center qubit in a odd chain, which accounts for the iSWAP-like operation between mirror-symmetric qubits.
Using local equivalence of graph states, we then obtain the star graph state (Supplementary Fig.~\ref{fig:mapping}\textbf{c}), composed of a central node connected to all others.
Here, a $Z_{-\pi/2}$ gate is applied on all qubits, except the qubit chosen to be the central node, where a $X_{\pi/2}$ gate is applied instead.
Finally, applying a Hadamard gate on all qubits except the central node transforms the star graph state into the GHZ state (Supplementary Fig.~\ref{fig:mapping}\textbf{d}).
Collectively, these transformations result in a final layer of single-qubit gates comprising $Y_{\pi/2}$ rotations on all qubits, except for: an arbitrary qubit in even chains, where instead a $Z_{-\pi/2}$ followed by an $X_{\pi/2}$ are applied; the centre qubit in odd chains, where instead a $X_{\pi/2}$ rotation is applied (choosing the central node of the star state to coincide with the centre qubit).

\begin{figure}[t]
    \includegraphics{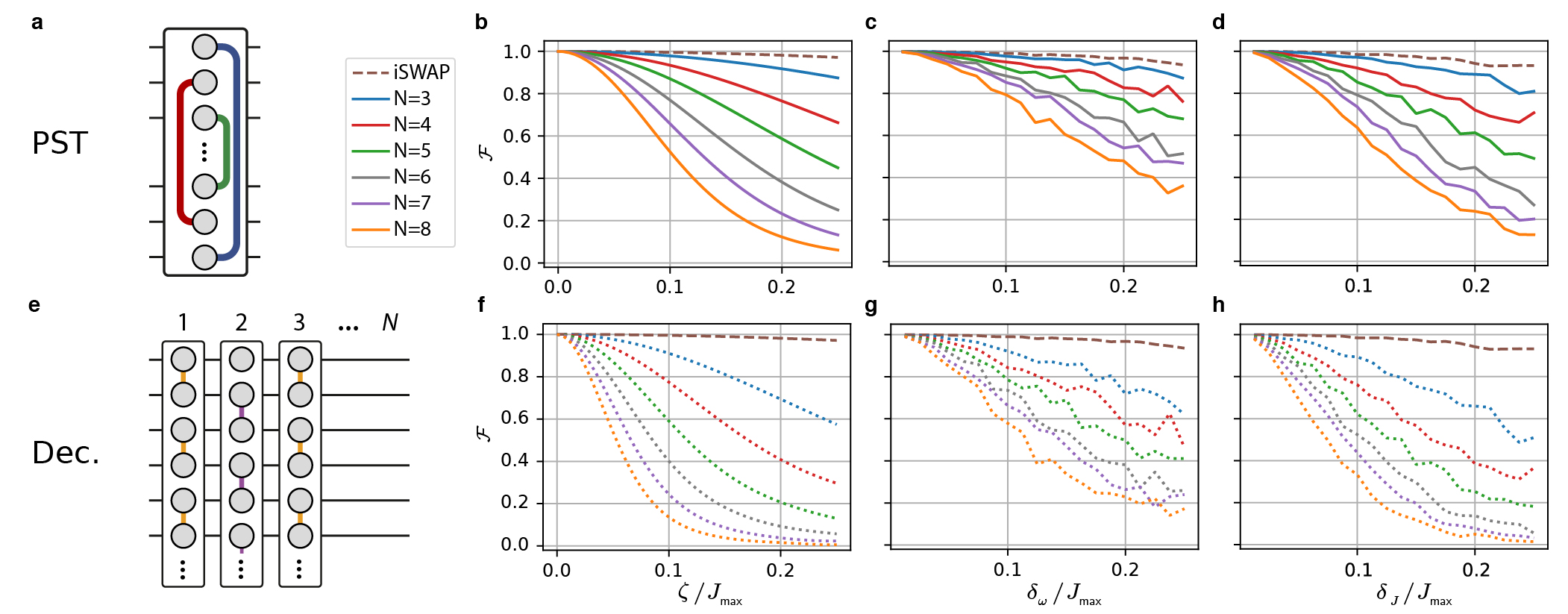}
    \caption{
    \label{fig:decomp}
    $\mybar$ \textbf{Error sensitivity of the PST operation and its equivalent decomposition.}
    Unitary fidelity overlap with ideal PST operation as a function of ZZ-type errors~\textbf{b},\textbf{f}, detuning errors~\textbf{c},\textbf{g} and coupling errors~\textbf{d},\textbf{h}.
    The top row shows the errors for PST (shown in~\textbf{a}) and the bottom row for its equivalent decomposition into the repeated application of two-qubit iSWAP gates (shown in~\textbf{e}).
    ZZ errors are applied with equal strength on the whole chain.
    Detuning and coupling errors are sampled from normal distributions with standard deviation $\delta_\omega^i$ and $\delta_J^i$, respectively. 
    100 realizations of the noise are averaged in the numerical simulation for each data point.
    }
\end{figure}

\subsection{-- Robustness of PST and optimal decomposition}
\label{sec:Robustness}
The decomposition of the PST into two-qubit gates operation discussed in the main text provides an intuitive understanding and a clear mapping to the graph state representation.
However, a more efficient decomposition exists based on the alternating application of iSWAP gates~\cite{Naegele2022}, shown in Supplementary Fig.~\ref{fig:decomp}\textbf{e}.
Derived in the context of Fermi-Hubbard simulations~\cite{Kivlichan2018, Cade2020}, this gate sequence is, to the best of our knowledge, the most time- and gate-efficient decomposition of the PST operation.
Nonetheless, PST is at least a factor of two faster than this efficient decomposition, as $\tau_\text{PST}\leq\frac{N\pi}{4J_\text{max}}$ for a chain length $N$, and the decomposition takes $\tau_\text{decomp}=N \times \tau_\text{iSWAP}=\frac{N\pi}{2J_\text{max}}$.
The increased speed of PST improves its robustness to relaxation and dephasing errors, since these scale linearly with time in the small error limit.
On the other hand, the effect of coherent errors on both implementations is non-trivial.
Therefore we perform simulations in the presence of additional error terms in the chain Hamiltonian
\begin{equation}
    \hat{H}_{\text{err}}/\hbar =  \sum_{i=1}^{N-1} \frac{\zeta}{4} (\mathds{1}_i\mathds{1}_{i+1}-\mathds{1}_i\sigma^z_{i+1}-\sigma^z_i\mathds{1}_{i+1}+\sigma^z_i\sigma^z_{i+1}) + \sum_{i=1}^{N}  \Delta_\omega^i \sigma^z_i + \sum_{i=1}^{N-1} \Delta_J^i (\hat{\sigma}_i^-\hat{\sigma}_{i+1}^+ +\hat{\sigma}_i^+\hat{\sigma}_{i+1}^-),
\end{equation}
where the errors $\Delta_\omega^i, \Delta_J^i$ are randomly sampled from normal distributions with standard deviations $\delta_\omega^i, \delta_J^i$.
We then calculate the fidelity of each implementation from the unitary overlap with the ideal operation $\mathcal{F} = |\text{Tr}(U_{\text{err}}U_{\text{ideal}}^\dag)/d|^2$, where $d=2^N$ is the dimension of the Hilbert space.
Unwanted ZZ interactions produce phase errors accumulating over time and, therefore, affect the PST operation less than its decomposition, as shown in Supplementary Fig.~\ref{fig:decomp}\textbf{b} and \textbf{f}.
PST also shows increased robustness to detuning and coupling errors, as shown in Supplementary Fig.~\ref{fig:decomp}\textbf{c-d} and \textbf{g-h}.
These results demonstrate that the PST operation is more robust than its decomposition to a variety of error sources in spite of the added complexity of the dynamics.

While the sequence discussed here is the exact decomposition of the PST operation, the gate-optimal sequence for the generation of GHZ states is a single Hadamard gate followed by CNOT gates to entangle each following qubit.
To minimise the total duration, the sequence can be started in the middle of the chain, and then, after the first CNOT, at each step, two CNOT gates can be applied simultaneously, one on each side.
The total operation time is then given by $\lceil N/2 \rceil \times \tau_\text{CNOT}$.
Assuming that the CNOT operation is implemented using a CZ gate and Hadamards, $\tau_\text{CNOT}>\sqrt{2} \tau_\text{iSWAP}$.
Then, as this sequence takes at least $\frac{N}{2}\times\tau_\text{CNOT}>\frac{N}{\sqrt{2}}\times\tau_\text{iSWAP}$, PST still provides a speed improvement of at least a factor of $\sqrt{2}$.
Hence, we expect similar fidelity improvements as described above.

\begin{figure}[t]
    \centering
    \includegraphics{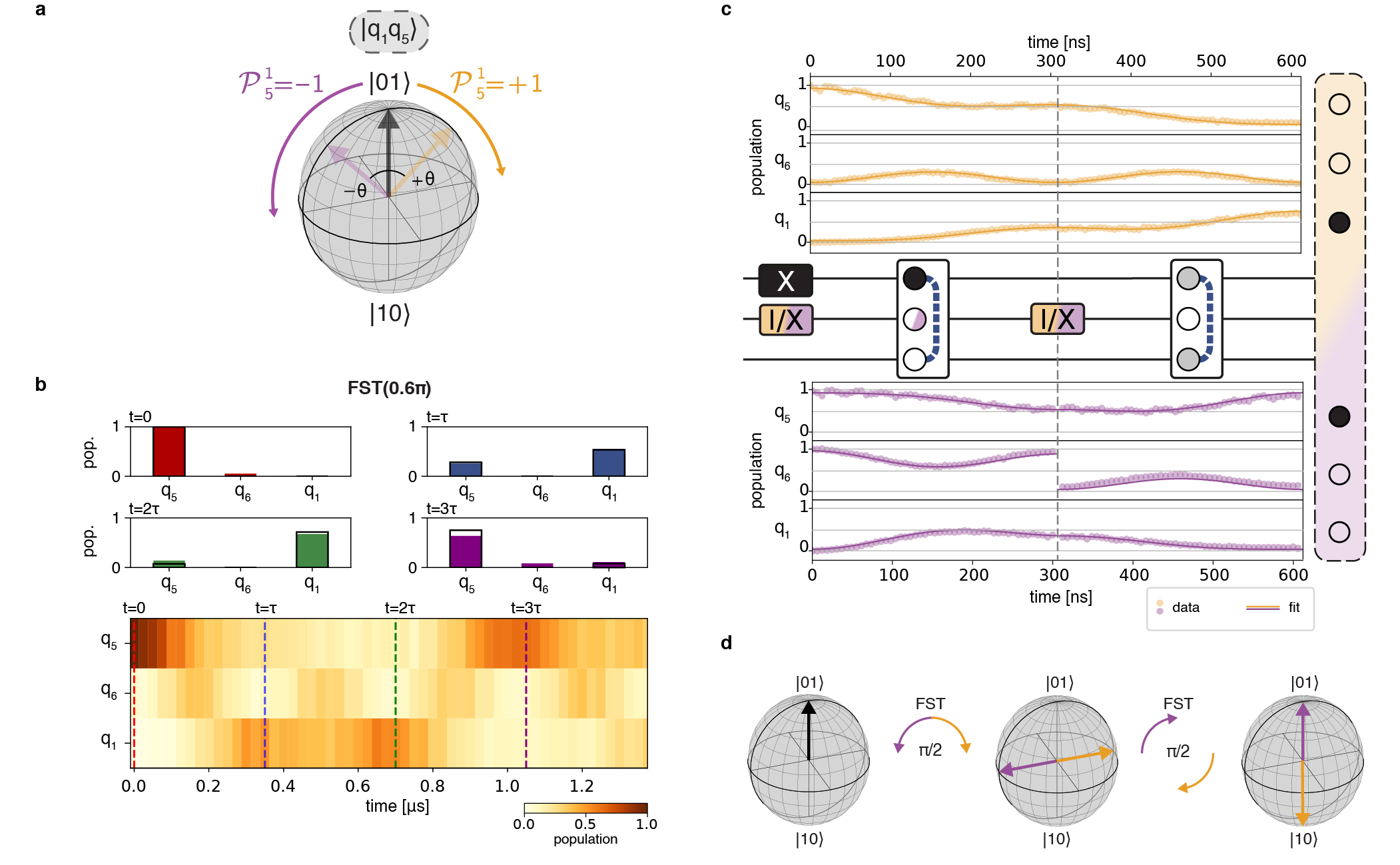}
    \caption{
    \label{fig:FST}
    $\mybar$ \textbf{Fractional state transfers on a three-qubit chain.}
    \textbf{a}~Bloch sphere visualisation of FST.
    The fractional transfer results in a rotation in the Bloch sphere spanned by the single-excitation states of the two outer qubits.
    The rotation angle $\pm\theta$ depends on the presence (orange) or absence (purple) of an excitation in the center qubit.
    \textbf{b}~Single-excitation dynamics of FST for $\theta=0.6\pi$. 
    The excitation is partially transferred between the two outer qubits after integer multiples of the transfer time $\tau = \SI{350}{ns}$.
    Contour plot (bottom) shows the dynamics, and solid-bar plots highlight the excited state populations of all qubits for $t=0, \tau, 2\tau, 3\tau$ (top).
    Black wireframes show the expected distribution renormalised by the experimentally observed decay.
    \textbf{c}~Parity-dependence on double-transfer dynamics.
    Qubits populations when preparing a single excitation on one of the outer qubits and applying $\text{FST}(\pi/2)$ twice.
    Measurement are repeated for the center qubit prepared in the ground state (orange plot and gates) or the excited state (purple plot and gates) during the first $\text{FST}(\pi/2)$. 
    Dots represent measured population and solid lines show simulated dynamics under the fitted chain Hamiltonian from Eq.~\eqref{eq:H_chainFST}. 
    \textbf{d}~Bloch sphere visualisation of the double-transfer experiments.
    State of qubits $q_1$ and $q_5$ are represented on Bloch sphere for different instances in the sequence:
    after state preparation (left), after the first $\text{FST}(\pi/2)$ operation (centre) and after the second one (right).
    Colors represent the parity during the first transfer.
    }
\end{figure}

\subsection{-- Fractional State Transfer}
\label{sec:FST}
Fractional State Transfer (FST) can be viewed as a generalised form of PST, granting partial transfer between mirror-symmetric qubits by an arbitrary amount.
In our system, FST is achieved by allowing the parametric drive frequencies $(\omega_{\text{d}_n})$ to be detuned from the difference frequency of the respective adjacent qubits $(\Delta_n)$.
Revisiting the procedure described in the main text yields the chain Hamiltonian \begin{equation}
    \label{eq:H_chainFST}
    \hat{H}_{\text{chain}} / \hbar = -\sum_{n=1}^{N}\frac{\delta_n}{2}\hat{\sigma}_n^z + \sum_{n=1}^{N-1}J_n(\hat{\sigma}_n^-\hat{\sigma}_{n+1}^+ + \text{h.c.}),
\end{equation}
where $(\delta_n)$ are the qubit frequencies in the drive's frame, satisfying $\delta_{n+1}-\delta_n = \Delta_n-\omega_{\text{d}_n}$.
A FST for an arbitrary transfer angle $\theta$ is obtained by setting the coupling strengths and qubit frequencies as
\begin{align}
    \hspace{10mm}
    \delta_n = 
    \begin{cases} 
    0 \\
    \frac{\pi}{2\tau} \left(\frac{\theta}{\pi} - 1\right) \frac{N}{2} \left(\frac{1}{2n-N} - \frac{1}{2n-2-N}\right)
    \end{cases}
    J_n = 
    \begin{cases} 
    \frac{\pi}{2\tau} \sqrt{\frac{n(N-n)((N-2n)^2 - (\frac{\theta}{\pi})^2)}{(N-1-2n)(N+1-2n)}} &\hfill \hspace{3mm} \text{for $N$ even,} \hspace{3mm}\\
    \frac{\pi}{2\tau} \sqrt{\frac{n(N-n)((N-2n)^2 - (\frac{\theta}{\pi}-1)^2)}{(N-2n)^2}} &\hfill \hspace{3mm} \text{for $N$ odd,} \hspace{3mm}
    \end{cases}
\end{align}
resulting in the coherent transfer of the fraction $\sin^2(\theta/2)$ of an excitation between any qubit and its mirror-symmetric counterpart.
Equivalently to PST, the FST protocol can also be described by an effective Hamiltonian~\cite{Naegele2022}
\begin{equation}
    \label{eq:H_FST}
    \hat{H}_{\text{FST}}/\hbar = \frac{\theta}{2\tau} \sum_{n=1}^{\lfloor\frac{N}{2}\rfloor}\bigg( \bigotimes_{k=n+1}^{\tilde{n}-1}\hat{\sigma}_k^z\bigg)(\hat{\sigma}_n^-\hat{\sigma}_{\tilde{n}}^+ + \text{h.c.}).
\end{equation}
whose dynamics are stroboscopically equivalent to the dynamics described by Eq.~\eqref{eq:H_chainFST}.
For every mirror-symmetric pair of qubits $\text{q}_n$ and $\text{q}_{\tilde{n}}$, the unitary evolution $U=e^{-i\hat{H}_{\text{FST}}\tau/\hbar}$ describes a rotation by a transfer angle of $\pm \theta$ in the Bloch sphere spanned by the two single-excitation states, as shown in Supplementary Fig.~\ref{fig:FST}\textbf{a}. The sign of the rotation angle depends on the number parity of excitations between $\text{q}_n$ and $\text{q}_{\tilde{n}}$, once again described by the operator $\hat{\mathcal{P}}_{n+1}^{\tilde{n}-1}=\bigotimes_{k=n+1}^{\tilde{n}-1}\hat{\sigma}^z_k \in \{-1, 1\}$.

We implement FST on a chain of $N=3$ qubits, $\text{q}_5$, $\text{q}_6$ and $\text{q}_1$, with a transfer time $\tau=\SI{350}{ns}$ and transfer angle $\theta=0.6\pi$, as shown in the dynamics of Supplementary Fig.~\ref{fig:FST}\textbf{b}. 
Solid-bar plots highlight the population on each qubit for multiples of the transfer time $\tau$.
The excitation only partially transfers between the outer qubits and matches the expected distributions, shown as black wireframes. 

In order to highlight the parity-dependence property of FST, we perform two experiments involving repeat applications of an FST with transfer angle $\theta=\pi/2$, shown in Supplementary Fig.~\ref{fig:FST}\textbf{c}-\textbf{d}.
In both cases, we prepare a full excitation on one of the outer qubits, $\text{q}_5$, followed by two identical $\text{FST}(\pi/2)$ operations and a measurement to all qubits.
In the second experiment, however, an extra excitation is initially prepared on the middle qubit $\text{q}_5$ and removed right before the second $\text{FST}(\pi/2)$ operation by an additional $X_{\pi}$ pulse.
Because $\langle \hat{\mathcal{P}}_{5}^{1}\rangle =1$ for both FST operations in the first sequence, the corresponding Bloch sphere rotations happen in the same direction and amount to a full population transfer to the mirror-symmetric qubit $\text{q}_1$.
In the second sequence, however, the two transfers produce rotations with opposite directions since $\hat{\mathcal{P}}_{5}^{1}$ changes from negative to positive, resulting in the excitation being refocused back to the original qubit, $\text{q}_5$.

\begin{figure}[t]
    \includegraphics{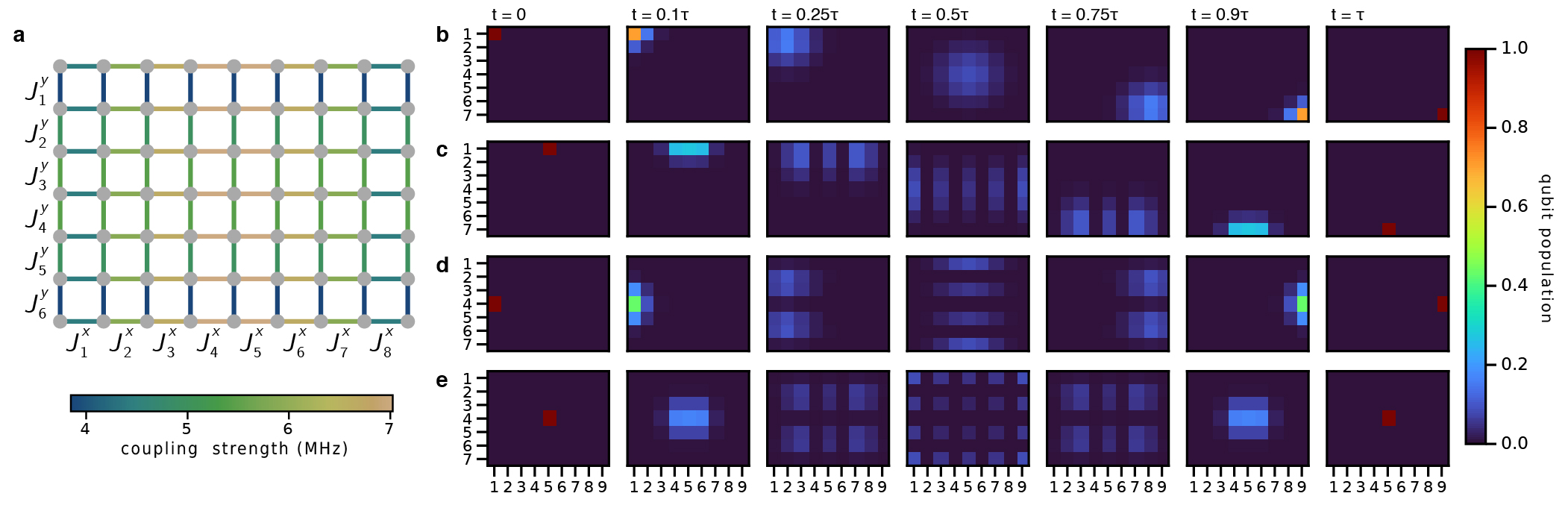}
    \caption{
    \label{fig:PST2D}
    $\mybar$ \textbf{Simulations of PST in a two-dimensional square lattice.}
    \textbf{a}~Two-dimensional $9$-by-$7$ qubit square lattice with couplings engineered for PST with a transfer time of $\tau=\SI{1}{\micro\second}$.
    Couplings strengths from position $(x=i,y)$ to $(x=i+1,y)$ are given by $J_i^x$ and from position $(x,y=j)$ to $(x,y=j+1)$ are given by $J_j^y$.
    \textbf{b}-\textbf{e}~Simulated dynamics of excitations started at different positions of the lattice, shown at times $t=0, 0.1\tau, 0.25\tau, 0.5\tau, 0.75\tau, 0.9\tau$ and $\tau$.
    }
\end{figure}

\subsection{-- PST on square lattices}
\label{sec:PST2D}
The simplest extension of qubit chains to higher dimensions is square lattices with arbitrary dimensions.
To achieve PST in these structures, the couplings are engineered so as to satisfy the PST formula $J_n=\frac{\pi}{2\tau}\sqrt{n(N-n)}$ for every axis direction with equivalent transfer times $\tau$~\cite{Kay2010}, as shown in Supplementary Fig.~\ref{fig:PST2D}\textbf{a} for a two-dimensional square lattice.
Here, $J_i^x$ and $J_j^y$ represent the coupling strengths between positions from column $i$ to column $i+1$ and from row $j$ to row $j+1$, respectively.
Due to this structure the two dimensions behave independently, with PST happening simultaneously on each dimension, as shown in simulation for different initial excitation locations in Supplementary Fig.~\ref{fig:PST2D}\textbf{b}-\textbf{e}.
Note that, since the Jordan-Wigner transformation does not apply in higher dimensions, here PST is no longer ensured when multiple excitations are present in the lattice.

\begin{landscape}
\begin{figure}[]
\centering
\includegraphics[width=247mm]{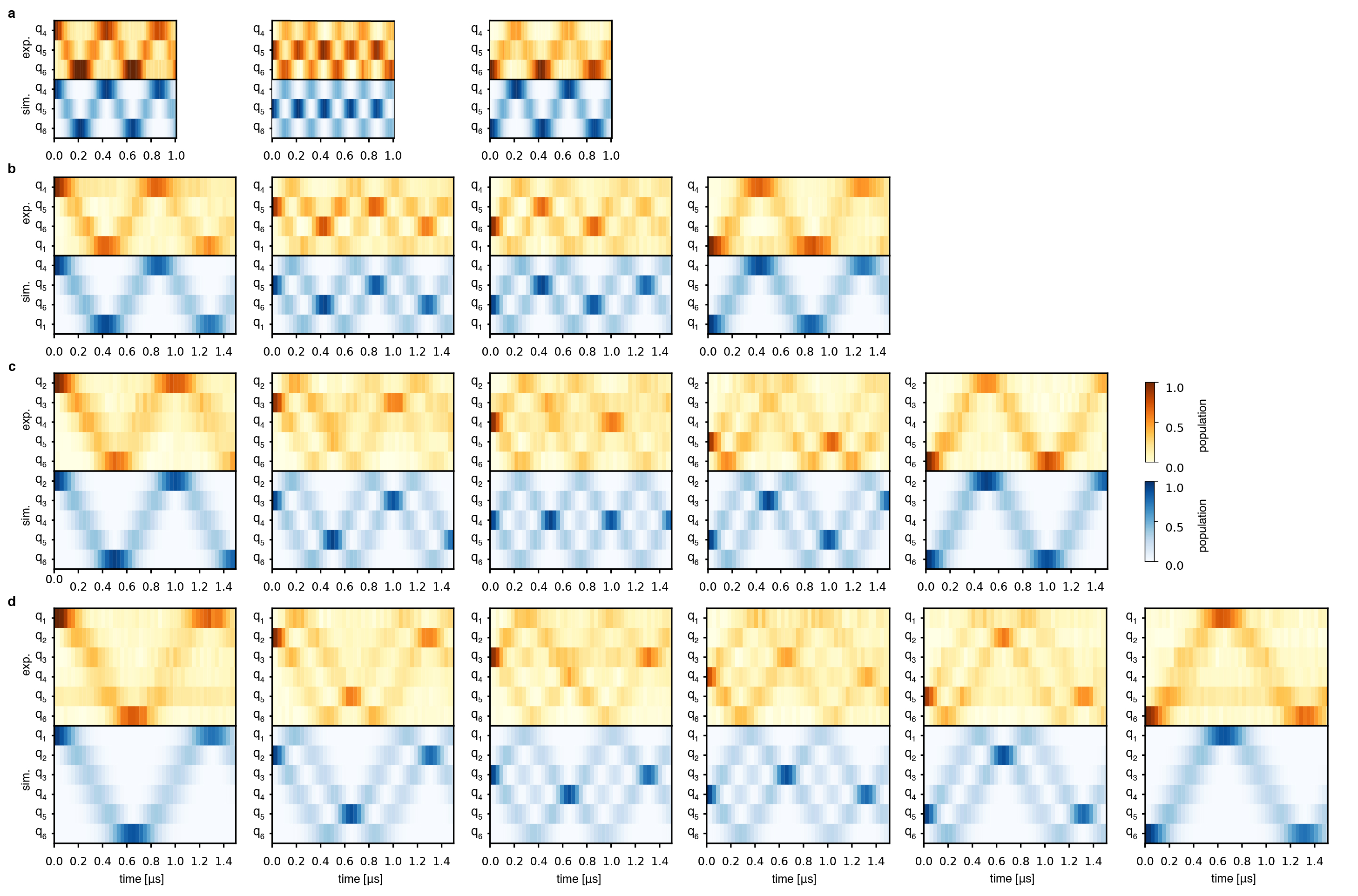}
\caption{
    \label{fig:all_PST}
    $\mybar$ \textbf{Perfect State Transfer protocol for various chain lengths.} 
    All single-excitation dynamics are shown for illustrative cases with three \textbf{a}, four \textbf{b}, five \textbf{c} and six \textbf{d} qubits.
    For each chain, experimental data (orange) is compared to simulation with ideal coupling strengths and relaxation effects for the respective qubit subset (blue). 
}
\end{figure}
\end{landscape}

\end{document}